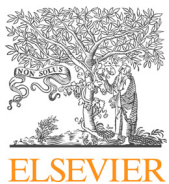
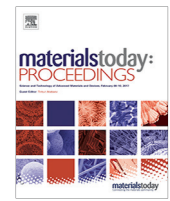

Contents lists available at ScienceDirect

# Materials Today: Proceedings

journal homepage: www.elsevier.com/locate/matpr

# A mathematical model of COVID-19 transmission

R. Jayatilaka [a], R. Patel [a], M. Brar [a], Y. Tang [a], N.M. Jisrawi [a,b], F. Chishtie [c,⁎], J. Drozd [d,⁎], S.R. Valluri [a,b,⁎]

[a] *Department of Physics and Astronomy, University of Western Ontario, 1151 Richmond Street, London N6A 3K7, Canada*
[b] *Mathematics Department, King's University College, University of Western Ontario (UWO), 266 Epworth Avenue, London N6A 2M3, Canada*
[c] *Department of Applied Mathematics, University of Western Ontario, Canada*
[d] *Mathematics Department, Huron University College, UWO, London N6G 1H3, Canada*

ABSTRACT

Disease transmission is studied through disciplines like epidemiology, applied mathematics, and statistics. Mathematical simulation models for transmission have implications in solving public and personal health challenges. The SIR model uses a compartmental approach including dynamic and nonlinear behavior of transmission through three factors: susceptible, infected, and removed (recovered and deceased) individuals. Using the Lambert W Function, we propose a framework to study solutions of the SIR model. This demonstrates the applications of COVID-19 transmission data to model the spread of a real-world disease. Different models of disease including the SIR, SIRmp and SEIRρqr model are compared with respect to their ability to predict disease spread. Physical distancing impacts and personal protection equipment use are discussed with relevance to the COVID-19 spread.

© 2021 Elsevier Ltd. All rights reserved.
Selection and peer-review under responsibility of the scientific committee of the International Conferences & Exhibition on Nanotechnologies, Organic Electronics & Nanomedicine ā NANOTEXNOLOGY 2020 This is an open access article under the CC BY license (http://creativecommons.org/licenses/by/4.0/).

## 1. Introduction

The First World War ravaged the world with death and destruction. A key contributor to the enormous death toll was not war, but a product of its chaotic environment; the 1918 "Spanish" Influenza. This H1N1 virus of avian origin spread throughout 1918–1919, infecting over 500 million individuals, and killing at least 40 million people worldwide [31,21]. Lack of sanitation and resources during wartime, and no progress in the development of a vaccine, limited worldwide control efforts to non-pharmaceutical interventions such as isolation and use of informal disinfectants [14]. Due to the immense, rapid spread of disease, countries were unable to suitably prepare themselves to prevent or control the influenza.

Now, almost a century later, the world is rocked again by the emergence of the new strand of coronavirus disease (COVID-19). This novel virus was first reported in December 2019 in Wuhan, China and has since spread to pandemic proportions [28]. As this virus can be transmitted person to person [28], many protective measures such as masking, social distancing, and most recently vaccines have been put in place to reduce the danger of human interactions.

COVID-19 targets the human respiratory system, resulting in clinical findings such as high fever, dyspnea and invasive multi-lobed lesions as seen in chest radiographs [18,27]. It has been reported that the symptoms of this virus start about 5 days after contracting it [28]. These symptoms tend to get progressively worse as time goes on, some cases leading to death, while others successfully recover [18]. This is a major public threat since thousands of Canadians have been hospitalized due to respiratory issues along with other flu-like symptoms after being diagnosed with COVID-19 [28]. Significant effort has been made in the last year to develop new and effective vaccines [41,42].

While the world now has the advantage of more accessible resources and a better understanding of pandemics compared to 1918, there are still the problems of disease prevention and control. A way to combat this is to model the disease over time, to better understand the gravity of the situation [5]. Epidemics play a major role in understanding disease transmission by studying disease distribution, sources of diseases, causes of diseases, and methods of disease control [16]. Using data of epidemic curves, one can use mathematical models to extrapolate disease data and trends to prepare for potential disease burden and determine public policies to mitigate risks of spread. Calvetti et al. have used *meta-*

⁎ Corresponding authors.
E-mail addresses: fachisht@uwo.ca (F. Chishtie), jdrozd1@uwo.ca (J. Drozd), valluri@uwo.ca (S.R. Valluri).

population network models for understanding, predicting, and managing the COVID-19 epidemic [3,5].

The Susceptible-Infected-Removed (SIR) model and its derivatives is one way to understand the transmission of diseases and predict future outcomes regarding COVID-19 cases. This study uses the SIR model, and variations of it such as the SIRmp model, Susceptible-Exposed-Infected-Removed (SEIR) model, and the SEIRρqr to illustrate COVID-19 spread. This study also uses the Lambert W function to analyze the SIR and SEIR models to better understand disease spread.

Section 2 of this paper discusses data and methods used to illustrate COVID-19 trends through different models such as the SIR, SIRmp, SEIR and the SEIRρqr models. By using Canadian data to model the current trend of COVID-19, it is possible to create graphs that depict where the individuals stand with respect to time during the spread of disease. Using a mixing factor m, it is possible to introduce a human-behaviour or social distancing factor into the situation. Section 3 presents the results of our work and simulations of Canadian COVID-19 data in context of the aforementioned models. In Section 4, the results obtained from our analysis are discussed. Finally, Section 5 of this paper presents our conclusions.

## 2. Methods

In this study, open-source COVID-19 datasets provided by Public Health Agency of Canada's Public Health Infobase is used [4]. The data ranges from January 22, 2020 to September 6, 2021, with each time series tracking an epidemiology statistic. The three-time series of focus are count of confirmed cases, deaths, and recovered cases nation-wide in Canada.

### 2.1. SIR model

The SIR model is a representation that divides a population with respect to a disease's impact on an individual over time. An individual can be categorized as susceptible ($S(t)$), infected ($I(t)$), or removed ($R(t)$, dead or cured), denoted by S, I and R respectively, along an independent variable, time [29]. One of the most common SIR models is the classic Kermack–McKendrick Model for contagious diseases in a closed population over time, which illustrates rapid changes in the number of infected patients during epidemics. It is assumed that there is a fixed homogeneous population size, random population mixing, instantaneous incubation period, and acute onset of disease [15,34,35]. The model variables can be represented as fractions:

$$s = \frac{S}{N} \tag{1}$$

where $s$ is a fractional representation of the number of susceptible individuals ($S$) over a selected population ($N$) over time.

$$i = \frac{I}{N} \tag{2}$$

where $i$ is a fractional representation of the number of infected individuals ($I$) over a selected population ($N$) over time.

$$r = \frac{R}{N} \tag{3}$$

and $r$ is a fractional representation of the number of removed individuals ($R$) which include the recovered and deceased individuals over a selected population ($N$) over time.

Overall, these equations must add to 1:

$$s + i + r = 1 \tag{4}$$

Using these equations, it is possible to extract three nonlinear differential equations that aid in tracking the illness progression. We present these equations below.

The Susceptible Equation:

$$\frac{ds}{dt} = -\beta si \tag{5}$$

where $\beta$ represents the infection rate, the probability per day that an I-person can infect a S-person, assuming the absence of social distancing. The Infected Equation:

$$\frac{di}{dt} = \beta si - \gamma i \tag{6}$$

The Removed Equation:

$$\frac{dr}{dt} = \gamma i \tag{7}$$

where the effective $\gamma$ represents the removal rate (encompassing both the recovered and deceased individuals), the probability per day that an I-person transitions into an R-person (becoming non-infectious permanently) [6].

The ratio of S-persons transitioning into I-persons is the ratio of $\beta$ to $\gamma$, referred to as the Reproduction Number; $\lambda$.

$$\lambda = \frac{\beta}{\gamma} \tag{8}$$

The higher the value of $\lambda$, the more transmittable the disease is; the infection rate eclipses the removal rate.

While $R_0$ usually denotes the reproduction number, this paper uses $R_0$ to denote the initial value of the removed variable at time $t = t_0$.

There is always some natural immunity, so it is reasonable to assume that $r_0$ is greater than 0. If the population has been partly vaccinated, the value of $r_0$ might even be 0.40 or more. Similarly, even without vaccination, a prior asymptomatic spread of the disease in the population may have resulted in $r_0$ being perhaps 15 or 20 percent of the population [12].

Some other variables can be introduced for the SIR model for convenience of comparison with information reported about the course of the epidemic. The total number of cases since the beginning of the epidemic is C. The initial value of the total number of cases, prior to time $t = t_0$, is $C_0$. The fraction of the number of new cases per day is $j$. The variable $j$ is defined by:

$$j = \frac{dC}{dt} \tag{9a}$$

Therefore, considering a closed population ($N \gg 1$) this equation becomes:

$$j = \beta si \tag{9b}$$

where $j$ is the number of cases per day in a closed population.

There is a possibility that some individuals may have been included in the R-group due to natural immunity or vaccination immunity, rather than as recovered cases. Therefore, by tracking the decline in S-persons, it is possible to track the increase in total cases, c, while excluding the individuals with immunity [12]. This indicates that $s$ can be used as an independent variable to find $i$ as a function of $s$:

$$i(s) = 1 - s - r(s) \tag{10a}$$

where $r(s)$ can be written as $r_0 - \frac{1}{\lambda} \ln\left(\frac{s}{s_0}\right)$.

$$i(s) = 1 - s - r_0 + \frac{1}{\lambda} \ln\left(\frac{s}{s_0}\right) \tag{10b}$$

Dividing Eq. (6) by Eq. (5) results in:

$$\frac{di}{ds} = \frac{\gamma - \beta s}{\beta s} \quad (11)$$

The solution of this equation for $i$ gives a Lambert W Function as implicitly seen in the expression given in Eq. (10b). This remarkable function has created a renaissance in the solution of diverse problems in innumerable fields of knowledge [9]. The solution is as follows:

$$s = -\frac{1}{\lambda} W(-\lambda c \, exp(\lambda i)) \quad (11a)$$

where $c = i_0 + s_0 - \frac{1}{\lambda}\ln(s_0)$ is the constant of integration to be determined from initial conditions by solving Eq. (11) with initial condition $i(s_0) = i_0$. Since Eq. (11a) has a Lambert W function with an exponential argument, this can also be expressed as an Omega Wright function [37].

To continue, it is possible to use $r$ as an independent variable as well. The expressions of $s$ can be found with respect to $r$:

$$s(r) = s_0 exp[-\lambda(r - r_0)] \quad (12a)$$

If an R-curve graph shows a continued increase, it would indicate an increase in number of removed individuals [3].

Eq. (12a) can then be substituted into the equation:

$$\frac{di}{dr} = \frac{\beta s - \gamma}{\gamma} = \lambda s - 1 \quad (12b)$$

to give:

$$\frac{di}{dr} = \lambda s_0 exp[-\lambda(r - r_0)] - 1 \quad (12c)$$

This Eq. (12c) can be integrated to provide an equation that illustrates $i$ as a function of $r$:

$$i(r) = i_0 + s_0\{1 - exp[-\lambda(r - r_0)]\} - (r - r_0) \quad (13a)$$

$$i(r) = 1 - r - s_0 exp[-\lambda(r - r_0)] \quad (13b)$$

If there are very few infectious people, the I-group becomes a very small fraction of the population, therefore $s + r \approx 1$. In addition, peak infections occur when $di/dt = 0$, the time when the I-group is the largest, assuming $t = t_1$ at $I_{max}$, it is possible to rework the Infection Equation as:

$$\beta s(t_1) i(t_1) = \gamma(t_1) i(t_1) \quad (14a)$$

$$\beta s(t_1) = \gamma \quad (14b)$$

$$s(t_1) = \frac{\gamma}{\beta} = \frac{1}{\lambda} \quad (14c)$$

Therefore, the lower the value of $\lambda$, the larger the number of people entering the R-group. This is as the removal rate will overpower the rate of individuals entering the I-group.

When $\lambda < 1$; $\gamma > \beta$. This indicates that the $s(t)$ curve will decrease past $r(t)$ curve, which will increase. When $\lambda = 1$, the ratios of $s(t)$ and $r(t)$ are equal and will inverse after the point of equivalence. When $\lambda < 1$, the ratio of $s(t)$ was greater than $r(t)$. This demonstrates that the $i(t)$ value was increasing as the infection rate, $\beta$, is greater than the removal rate, $\gamma$. A point of inflection occurs in the $i(t)$ curve at $I_{max} = t_1$ which illustrates that as the ratios inverse between the $s(t)$ and $r(t)$ curves. The $\lambda$ value decreases, indicating a lower $\beta$ value; implying a decrease in members in the I-group and a descending $i(t)$ curve.

The value of the inflection points can be found using the second derivative of $s$ with respects to $t$:

$$\frac{d^2 s}{dt^2} = -\beta \frac{d}{dt}[si] \quad (15)$$

As the epidemic dies out, the number of infectious people approaches zero, so an asymptotic limit is formed; $t \to \infty$, and therefore, $s + r = 1$. Inflection points will be discussed in greater detail in Section 4.2. It is interesting to observe that studies in the context of the inverse SIR model have been made by many researchers [6,8,38,40] who derive an analytic solution for the more general case of a time-dependent infection rate that is not limited to a certain range of parameter values. Kröger and Schlickeiser were able to relate all parameters of the SIR model to the cumulated number of infected population and its first and second derivatives at an initial time $t = 0$, where data is assumed to be available [38,40]. They could relate by a suitable dimensionless time variable tau to the natural logarithm of the susceptibility function $S(\tau)$ and thereby obtain an analytic solution for arbitrary time dependence of the infection rate. They derived expressions in terms of the Lambert-W function for $S(\tau)$ and also $j(\tau)$.

*2.1.1. SIRmp model*

The SIRmp model, as derived from the SIR model, focuses on the relationship between disease transmission and the effect of public health measures. Consider a situation in which public health guidelines are introduced to slow the frequency, duration and contact distance between S-people and I-people. This can be represented by making the value of the parameter $\beta$ and $\gamma$ vary with time with $m$ and $p$ as logistic equations as listed above Table 2 and are used as scaling coefficients of $\beta$ and $\gamma$. However, a conceptually simpler way to describe such public health measures is to keep $\beta$ constant and multiply it by a time-varying mixing factor $m$ to reflect changes in social distancing. In the present section, we assume that $\beta$ is constant, and develop the equations and approximations for the standard SIR model by setting all $m$ values to 1.

As such, the differential equations in population fraction notation are:

$$\frac{ds}{dt} = -\beta m s i \quad (16)$$

$$\frac{di}{dt} = \beta m s i - \gamma i \quad (17)$$

$$\frac{dr}{dt} = \gamma i \quad (18)$$

$$s + i + r = 1 \quad (19)$$

Dividing Eq. (17) by Eq. (16) results in:

$$\frac{di}{ds} = \frac{-\beta m s + \gamma}{\beta m s} \quad (20a)$$

The solution of equation of (20a) is given as:

$$s = -\frac{1}{\lambda m} W(-c\lambda m \, exp(\lambda m i)) \quad (20b)$$

where $c = i_0 + s_0 - \frac{1}{\lambda m}\ln(s_0)$ is the constant of integration to be determined from initial conditions by solving Eq. (20a), with initial condition $i(s_0) = i_0$.

This equation is in terms of the Lambert W function, which is defined after Eq. (27) below.

The equations for the fraction of total cases per day, $j$ and the total cumulative cases, $C$ are [33]:

$$j = \frac{dC}{dt} \quad (21a)$$

or

$$j = \beta m s i \quad (21b)$$

and

$$C = C_0 + \int_0^t j(\tau)d\tau \quad (22a)$$

or

$$C = C_0 + \beta \int_0^t m(\tau)s(\tau)i(\tau)d\tau \quad (22b)$$

### 2.2. SEIR model

The SIR model can be extended using the Susceptible-Exposed-Infected-Removed (SEIR) variant. This model also considers the susceptible, infected and removed populations but unlike the SIR model it also considers the exposed population; those who are incubating the virus but are not infectious or infected [22]. The SEIR model adds another layer of complexity to the SIR model, by allowing the analysis of conditions of susceptible and infected populations during an epidemic outbreak [11].

The SEIR model's governing equations are:

$$\frac{ds}{dt} = -\rho\beta si \quad (23)$$

$$\frac{de}{dt} = \rho\beta si - \alpha e \quad (24)$$

$$\frac{di}{dt} = \alpha e - \gamma i \quad (25)$$

$$\frac{dr}{dt} = \gamma i \quad (26)$$

where the parameters are defined as [5,25]:
$\alpha$: incubation rate from the exposed group to the infected group,
$\beta$: infection rate,
$\gamma$: removal rate from the infected group to removed group,
$\rho$: the reduced spread rate factor ($0 \leq \rho < 1$).
The equations have been modified to properly reflect a closed population.

This study examines the use of the Lambert W function in conjunction with the SIR and SEIR models, the multivalued inverse of the function $w \rightarrow we^w$ [9]. In the 18th century, scientist Johann Lambert gave a solution to a trinomial equation, upon which further work by Euler and Sir Edward Wright led to the now modern definition of Lambert's original work [32]. Their function, named to honour Lambert, is as follows:

$$W(z)e^{W(z)} = z \quad (27)$$

The Lambert W function is implicitly elementary in that it is defined by an equation composed of only elementary functions but is not an elementary function itself. It has applications in a variety of fields ranging from quantum physics, black holes to even the spread of disease [13].

Corless et al.'s article regarding the Lambert W function further studied the function's applicability in epidemics. Let us assume in a population of $n$ people, everyone has the same contact with $\alpha$ random others [9]. If $\gamma$ is the weak connectivity of this random net, and disease spreads through transitivity to those in close contact with the infected individual, the total infected population is approximated as $\gamma n$ for large $n$, where:

$$\gamma = 1 - e^{-\alpha\gamma} \quad (28)$$

This formula can also be applied for conditions where $\alpha$ is a fixed integer, as well as when $\alpha$ is an expected value in that it is not fixed for all individuals and may not be an integer [17,30]. Re-writing the above formula we obtain the following:

$$\alpha e^\alpha = \alpha(1 - \gamma)e^{\alpha(\gamma-1)} \quad (29a)$$

One can determine:

$$\gamma = 1 - T\frac{(\alpha e^{-\alpha})}{\alpha} = 1 - W\frac{(-\alpha e^{-\alpha})}{\alpha} \quad (29b)$$

where $\alpha \geq 1$, using the principal branch of T (of the Tree function) and W (of the Lambert W function) [9].

This epidemic problem is closely tied to a phenomenon described by Erdös and Rényi in which the epidemic problem is related to the size of the 'giant component' in a random graph [10]. Essentially, when a graph on $n$ vertices with $m = 1/2\, \alpha n$ edges is randomly chosen, it is almost certain it has a connected component with approximately $\gamma n$ vertices (for $\gamma$ given by Eq. (29b)) when $\alpha \geq 1$ [9].

#### 2.2.1. SEIRρqr model

The SEIRm model is a modification of the SEIR model, to introduce a mixing coefficient. The SEIRm model trials demonstrate various stabilities of the COVID-19 virus situation, based on an unpublished report and private communications by Ken Roberts [26]. By observing the value of $m$, the effect of social distancing policies and other public health measures can be estimated. An $m$ value of 1.0 represents normal (pre-pandemic) social mixing practices, and an $m$ value lower than 1 (say 0.6 or 0.8) represents the introduction of social distancing and other measures to reduce infectivity. Very low $m$ values are unrealistic, because of the economic impact. The ideas of a mixing factor are consistent with more recent work, involving mobility data and masking behavior. The paper of Comunian, et al(2020), is relevant to the SEIRm model. We further generalized the SEIRm model and modelled a SEIRρqr model with $\rho$, q, r being mixing coefficients represented by logistic equations as listed above Table 3 and are used as scaling coefficients for $\alpha$, $\beta$ and $\gamma$.

#### 2.2.2. The Planck-like black body distribution

While analyzing several SIR models of disease, it was observed that some of the infection curves looked like Planck's blackbody distribution curves due to the realistic asymmetry of the infection data curves [23]. Keeping this in perspective, it was decided this study would simulate infection curves using an asymmetric function rather than a purely symmetric one. Max Planck theorized that mode energies of the blackbody are not continuously distributed but are quantized. He devised a law for blackbody radiation as follows [2]:

$$B_\nu(T) = A\frac{(2h\nu^\alpha/c^2)}{e^{h\nu/kT} - 1}, \; \alpha = 3 \quad (30)$$

where the parameters are defined as:
$B_\nu$ : spectral radiance,
$h$: Planck's constant,
$c$: speed of light in a vacuum,
$k$: Boltzmann constant,
$\nu$: frequency of the electromagnetic radiation,
$T$: absolute temperature of the body,
$\alpha$: any value other than 3 to run Planck-like simulations in other situations.

Therefore, this formula represents the spectral-energy distribution of radiation emitted by a blackbody.

The similarity of the SIR model infection curve suggests that it may be reasonable to model the infection curve for a few different values of $\alpha$ like in a Planck Blackbody Distribution function with an appropriate definition of the constants $C_1$ and $C_2$ [32].

In this paper, two adjusted formulas inspired from the Planck-like Blackbody Distribution are proposed to model infection as a function of time.

$$I(t) = \frac{(C_2 \, t^\alpha)}{e^{C_1 \, t} - 1} \quad (31)$$

where $\alpha$ can be any positive integer.

## 3. Results

The figures below display the results of the all models (SIR, SIRmp, SEIRρqr) fitted onto the given Canadian COVID-19 dataset by parametrically solving the system numerically using ParametricNDSolve from Wolfram Mathematica (version 12.2.0.0). The respective model parameters were derived by using NonlinearModelFit to fit the data to $\beta$ and $\gamma$ for the first 320 days. For the sake of conciseness, the authors have focused on the first wave (subdivided into two time regions: 50–100 and 101–177 days) and the second wave (200–320 days). There is also a plot of the overall COVID removed and infected case counts (see Fig. 1) from where these days for each wave was taken from. *In* Figs. 2–9, *solid print lines refer to predicted trends, while the dotted lines refer to the Canadian data.*

During this study, it was found that none of the models were able to fully capture disease spread using one general approach – as such, it was found that parameters for the first wave had to be fitted separately for two separate time windows for the SIR and SEIR models as recommended in [7] and varied for SIRmp model in order to best capture disease spread. The ParametricNDSolve function in Mathematica 12.2.0.0 was used to solve the parameters. This is a proprietary Mathematica function that finds a numerical solution to a system of ordinary differential equations for a function with parameters. It solves the differential equations by going through several different stages, depending on the type of equations. Each stage is handled by a method. The actual stages used and their order are determined by the Mathematica function NDSolve based on the problem to be solved using boundary value problem methods. These functions and methods are proprietary within the Mathematica software [25]. The primary function of ParametricNDSolve and ParametricNDSolveValue is to process the input differential equations along with the parameters and return a ParametricFunction. When a specific set of parameters are provided, the ParametricFunction internally calls NDSolve to solve the problem. The implementation details of how NDSolve solves a particular problem is outlined in https://reference.wolfram.com/language/tutorial/NDSolveOverview.html.

In the tables, the P-value signifies the probability of finding the modeled results least extreme to the observations under the assumption of the null hypothesis. Hence, the smaller the p-value is, the less likely it is to violate the null hypothesis and the result is deemed significant. The t-statistic is the ratio of the departure of the estimated value of a parameter from observations to its standard error. It is generally the case that when these values are greater than 2 or less than −2, the model fit is better.

### 3.1. SIR model

For the SIR model, $\beta$ is estimated to be 0.18, 0.030 and 0.13 for the respective time intervals while $\gamma$ is estimated to be 0.070, 0.029, and 0.089 (refer to Table 1). The population (N) considered is 3,759,000, and initial infection, $i_0 = 1$ /N. The SIR model predicted the infected and removed case counts accurately for the first 55 days. After which, the predicted trends fail to capture the rise of infected and removed case counts as fast as they had occurred. Until day 94 in the infected curve and day 96 in the removed curve, the predicted trend was underestimated compared to the actual trend (Fig. 2). From days 101–177, the predicted trends linearly trace the infected and removed case counts (Fig. 3). In the removed curve from day 200–310 the predicted trend is severely underestimated compared to the actual trend. After day 310, the predicted trend is overestimated compared to the actual trend. The infected curve is accurately predicated until day 255. Day 256 onwards, the predicted trend is overestimated compared to the actual trend (Fig. 4).

#### 3.1.1. SIRmp model

For the SIRmp model, $\beta$ is estimated to be 0.18, 0.030 and 0.13 for the respective time intervals while $\gamma$ is estimated to be 0.070, 0.029, and 0.089 (refer to Table 2). The $m$ value used is 1.05 to modify $\beta$ and $p(t) = 1 - 0.004t$ as a variation in $\gamma$ (refer to Table 1). The variation of the SIR model parameters as a function of time was recently recommended in [38,40] as well. For the infected curve from days 56–94 the predicted curve underestimates the actual trend. From day 95 onward, the predicted trend overestimates the actual trend. In the removed curve, from days 70–100 the predicted trend is underestimated compared to the actual trend (Fig. 5). In the removed curve, from days 200–309 the predicted trend is underestimated compared to the actual trend. After day 310, the predicted trend is overestimated compared to the actual trend. For the infected curve, the trend is followed from days 200–255. From day 256 and onward, the predicted trend is overestimated compared to the actual trend (Fig. 6).

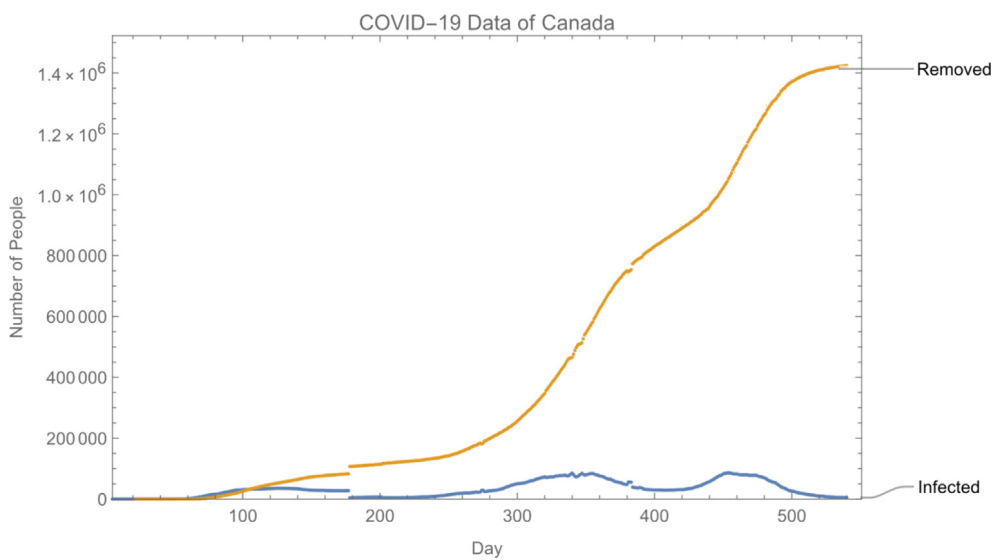

**Fig. 1.** The total number of infected (blue) and removed (orange) individuals in Canada. (For interpretation of the references to colour in this figure legend, the reader is referred to the web version of this article.)

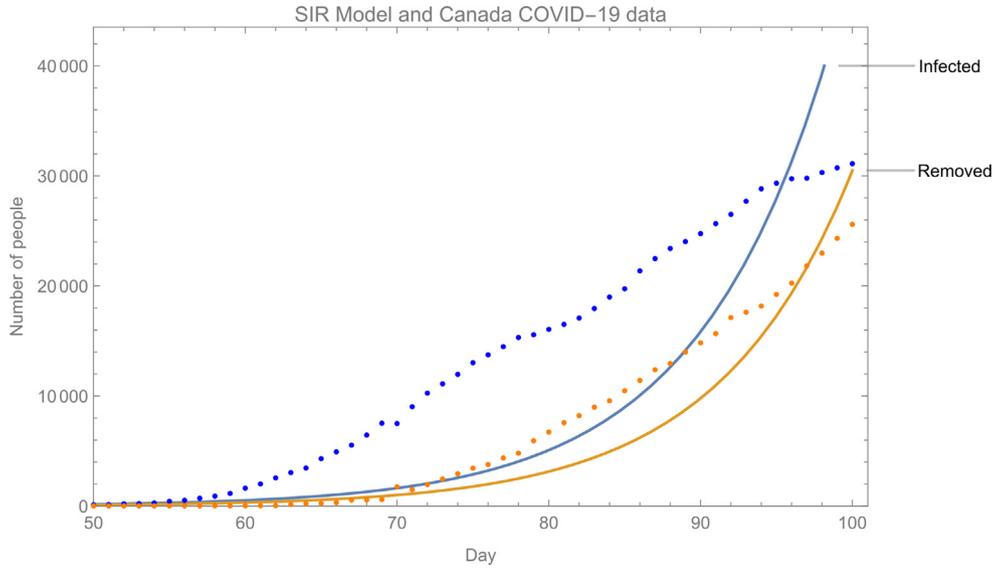

**Fig. 2.** SIR model prediction of infected and removed cases with respect to data for Canada for days 50–100.

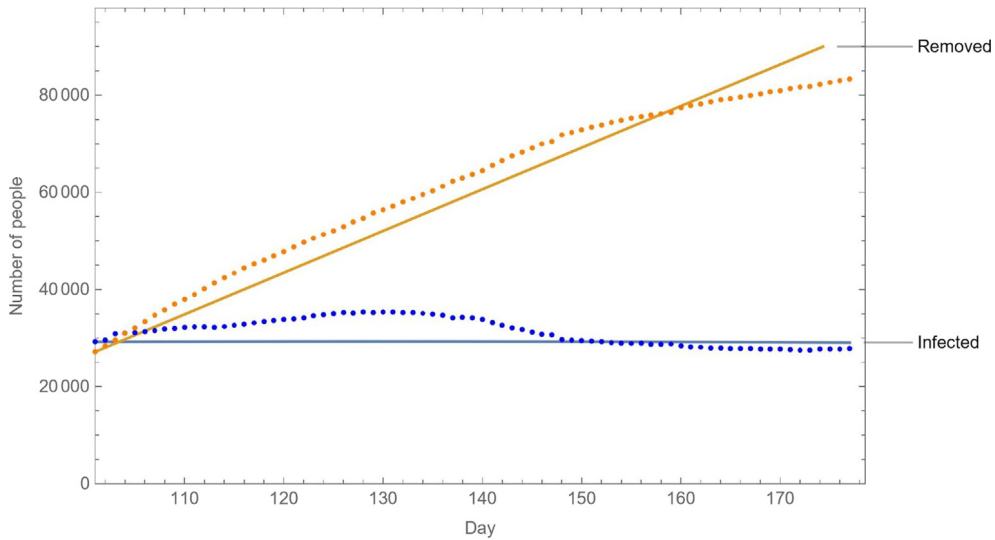

**Fig. 3.** SIR model prediction of infected and removed cases with respect to data for Canada for days 101–177.

Using the following logistic equations modifies the values of $m$ and $p$ to improve the overall SIRmp model performance:

$$m = \frac{1}{[1+e^{(-t)}]} \quad (32a)$$

$$p = \frac{0.7}{[1+e^{(-t)}]} \quad (32b)$$

### 3.2. SEIRρqr model

For the SEIRρqr model, it is assumed that $\rho$ = 1, q = 1 and r = 1 to produce the classic SEIR model results. In this model, $\alpha$ is estimated to be 0.030, and 0.19, $\beta$ is estimated to be 0.28, and 0.14 and $\gamma$ is estimated to be 0.0045, and 0.15 (refer to Table 3). Note that $\rho$ = 0 implies everyone in the society is quarantined, while $\rho$ = 1 implies no social distancing. The SEIRρqr model is able to follow the general trend of the actual case counts for days 50–100 of the first wave. In the infected curve, it is important to note that from days 69–92 the actual trend overestimates the predicted trend. From day 93 onwards the actual trend underestimates the predicted trend. The predicted trend for removed cases is slightly overestimated until day 80 (Fig. 7). In the second wave, from days 200–320, the predicted trends for both infected and removed follow the general linear trends. However, in the removed curve, it is important to note that from days 200–255 the predicted trend underestimates the actual trend. Also, in the infected curve, from day 285 onwards the actual trend overestimates the predicted trend (Fig. 8).

Using the following logistic equations modifies the parameters below and improves the SEIR ρqr model performance:

$$\rho = \frac{1}{[1+e^{(-t)}]} \quad (33a)$$

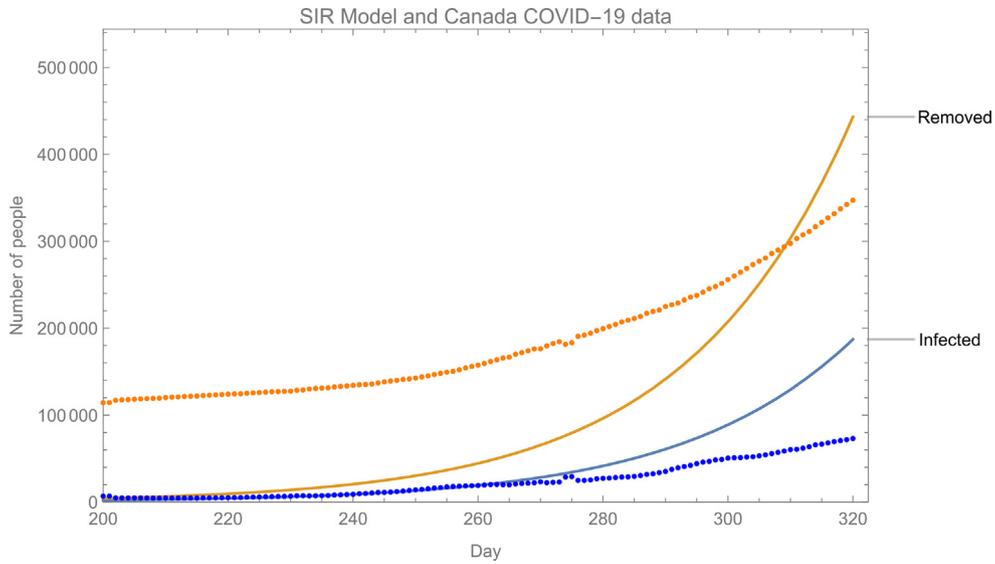

**Fig. 4.** SIR model prediction of infected and removed cases with respect to data for Canada for days 200–320.

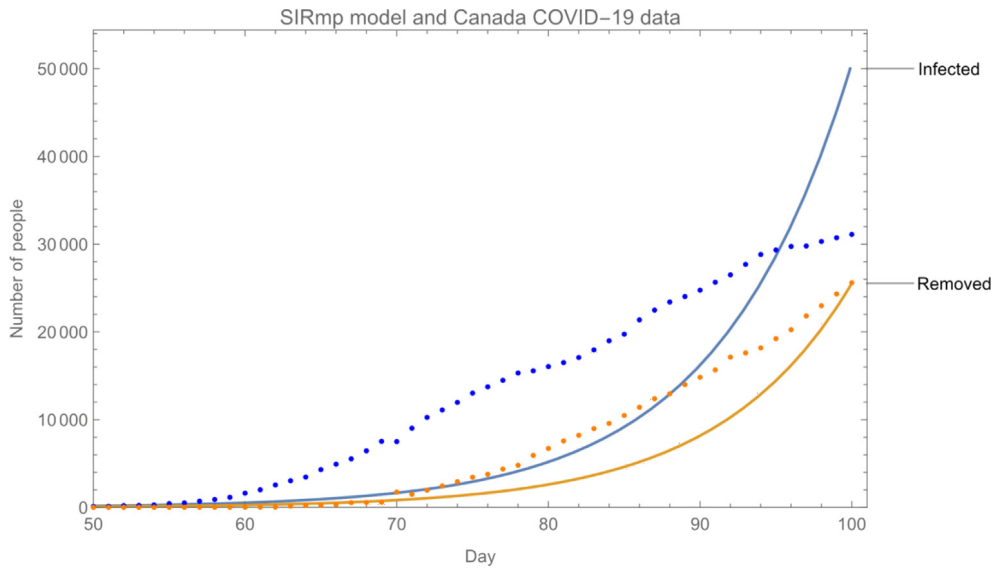

**Fig. 5.** SIRmp model prediction of infected and removed cases with respect to data for Canada for days 50–100.

$$r = \frac{0.8}{[1 + e^{(-t)}]} \quad (33b)$$

$$q = \frac{1}{[1 + e^{(-t)}]} \quad (33c)$$

### 3.3. Planck-Like blackbody function

Fig. 9 displays the results of the predicted infected curves after conducting a non-linear fit of the parameters $C_1$ and $C_2$. The parameters $C_1$ and $C_2$ were estimated to be 0.070 and $2.78 \times 10^{-11}$ respectively (see Table 4). Several trials of different $\alpha$ values were run, and it was determined that an $\alpha$ value of 9 yielded the best fit for modelling COVID-19 data (Fig. 9) getting a good estimation of the peak number of infected cases.

### 4. Discussion

#### 4.1. Model interpretations

Social distancing is the practice of reducing physical contact to reduce opportunity for spread of transmissible diseases [7]. Common practices include social isolation, self-quarantine and cancellation of mass gatherings. Matrajt and Leung used a mathematical model to illustrate how implementing social distancing measures earlier in an epidemic will delay the epidemic curve while interventions started later will flatten the curve. The model also illustrated that the epidemic would rebound when interventions are suspended, indicating the importance of maintaining social distancing practices for the safety of the population [20].

In this study, the SIR and SIRmp models demonstrate that while initially a good fit for modelling disease spread, it veers away from actual data as time passes since it fails to account for several anthropological factors such as adherence to prevention methods. The implication that $\beta$ and $\gamma$ values vary in the model to best fit the results suggest that models that vary these parameters would better fit the actual data, and therefore be able to better predict the disease spread. Overall, the SEIRρqr model was able to predict disease trends better, but it also fails to fully capture the impact of anthropological factors. These mathematical models offer predictions of disease trends in controlled situations that do not consider the impact of anthropological factors such as social distancing [8]. Comunian et al. (2020) have done a thorough study about the application of an inversion of the SIR model to the COVID-19 pandemic [8]. They have emphasized the importance of calibration of a SIR model with official international data for the COVID-19 pandemic. They discuss the inherent difficulties in the solution of inverse problems. The role and relevance of reliable data to provide proper calibration of the parameters is essential for successful model predictions. Calvetti et al. (2020) have used modified SEIR models that include infectious and asymptomatic individuals [3]. The network model uses easily interpretable parameters estimated from the available community data. They estimate their parameters using Bayesian techniques [3]. One way to combat this problem would be to focus on models that incorporate the addition of other factors such as public compliance and mixing factor. In recent work, the SEIRm model results displayed that the mixing factor, $m$, decreased rapidly to 0.2 levels over approximately the first 150 days since April 10, 2020 [2]. The $m$ factor then proceeds to increase to 0.3779 by September 16, 2020 [26].

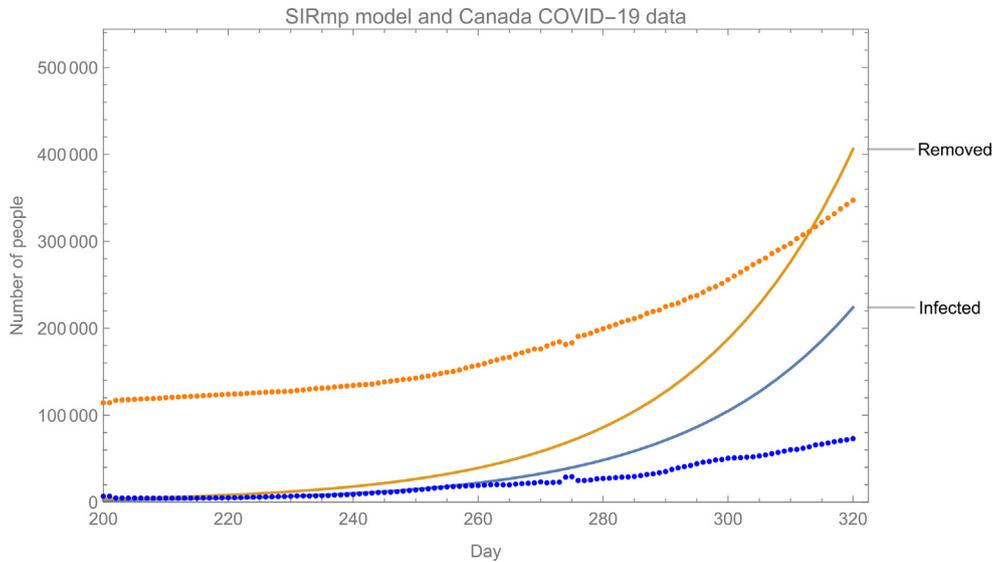

**Fig. 6.** SIRmp model prediction of infected and removed cases with respect to data for Canada for days 200–320.

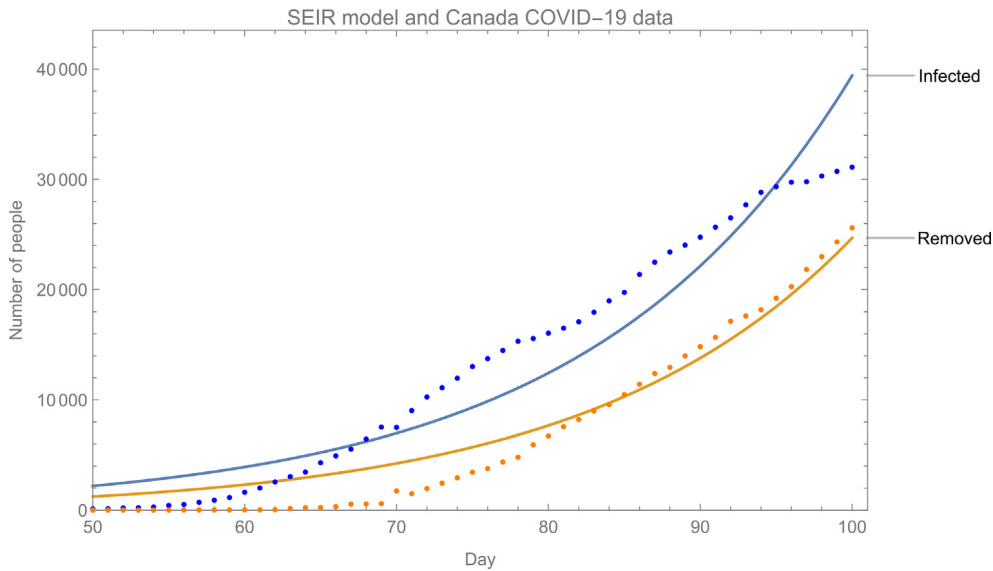

**Fig. 7.** SEIRρqr model prediction of infected and removed cases with respect to data for Canada for days 50–100.

The $m$ factor in the SEIRm model plays a crucial part in the significance of this model. The values of the m factor indicate the severity of the situation regarding COVID-19 case numbers. As indicated earlier, the greater the value of $m$, the more severe the situation. Regarding COVID-19, if a greater $m$ value was seen, this would indicate that numbers are rising which then puts greater pressure on hospitals due to a rapid increase in patients. A higher $m$ value would not only affect hospitals but would also impact equipment manufacturing companies and companies that are working to develop a vaccine for COVID-19. Alternatively, a lower $m$ value would indicate a more controlled or lower number of COVID-19 cases. This lessens the strain on hospitals, personal protection equipment manufacturers, and labs working on vaccine development. Moreover, the *m* value also allows for a hypothetical timeline to be developed. A timeline would be a very useful aid in creating a plan for various areas in order to properly control the spread of COVID-19.

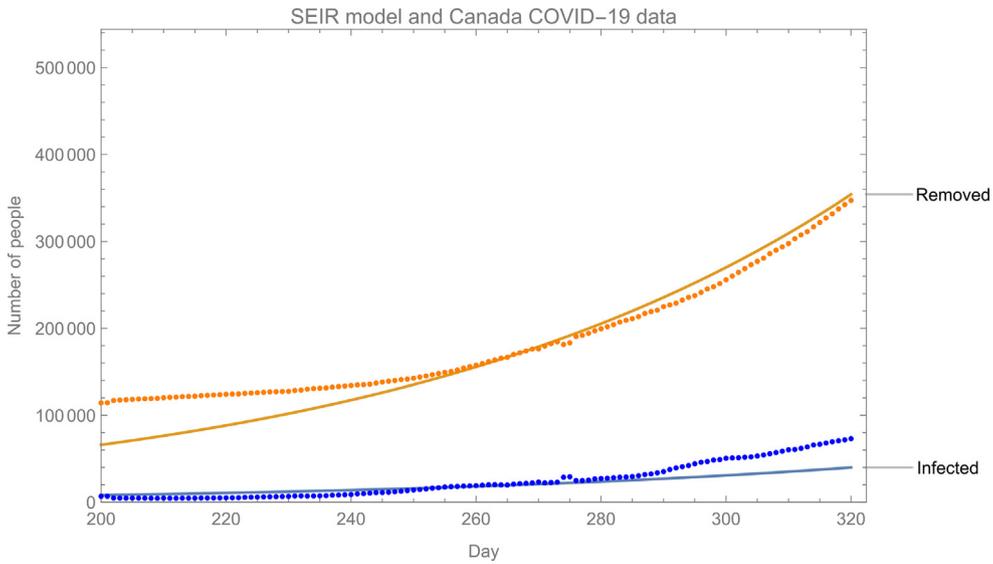

**Fig.8.** SEIRρqr model prediction of infected and removed cases with respect to data for Canada for days 200–320.

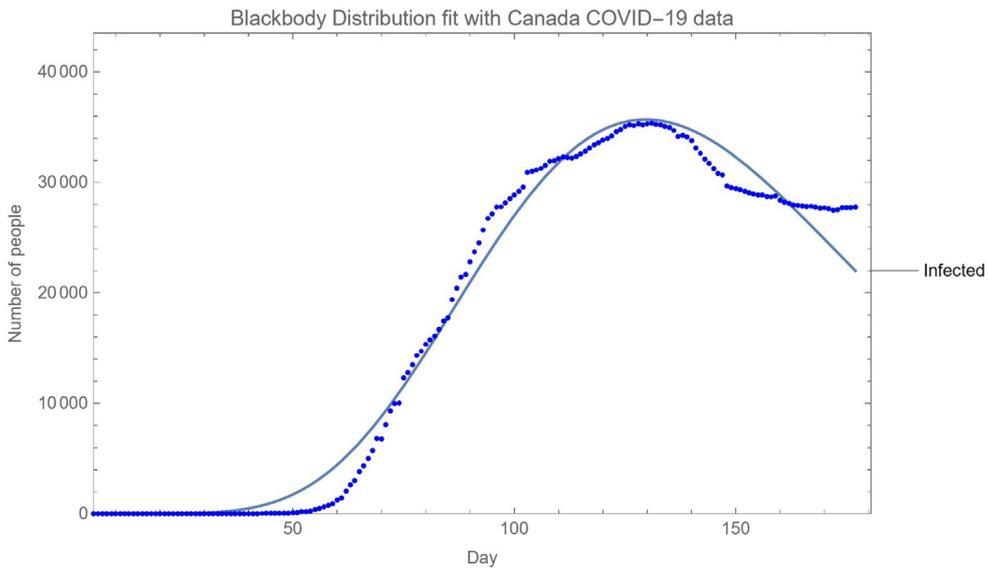

**Fig. 9.** Blackbody distribution fit of infected cases using a Planck Function ($\alpha = 9$) with respect to data in Canada from days 0–177.

**Table 1**
SIR model parameters used in Figs. 2-4.

| Days | Parameter | Estimate | Standard Error | t-Statistic | P-Value |
| --- | --- | --- | --- | --- | --- |
| 50–100 | $\beta$ | 0.18 | 0.0067 | 27.41 | $2.61 \times 10^{-48}$ |
| | $\gamma$ | 0.070 | 0.0070 | 10.07 | $6.97 \times 10^{-17}$ |
| 101–177 | $\beta$ | 0.030 | 0.00034 | 86.69 | $2.54 \times 10^{-131}$ |
| | $\gamma$ | 0.029 | 0.00043 | 68.04 | $9.82 \times 10^{-116}$ |
| 200–320 | $\beta$ | 0.13 | 0.00058 | 220.12 | $6.96 \times 10^{-279}$ |
| | $\gamma$ | 0.089 | 0.00057 | 156.46 | $1.49 \times 10^{-243}$ |

**Table 2**
SIRmp model parameters used in Figs. 5-6.

| Days | Parameter | Estimate | Standard Error | t-Statistic | P-Value |
| --- | --- | --- | --- | --- | --- |
| 50–100 | $\beta$ | 0.17 | 0.00059 | 292.71 | $1.71 \times 10^{-148}$ |
| | $\gamma$ | 0.082 | 0.00080 | 102.33 | $4.98 \times 10^{-103}$ |
| 200–320 | $\beta$ | 0.11 | 0.0068 | 15.79 | $5.45 \times 10^{-39}$ |
| | $\gamma$ | 0.099 | 0.010 | 9.71 | $5.01 \times 10^{-19}$ |

**Table 3**
SEIRρqr model parameters used in Figs. 7-8.

| Days | Parameter | Estimate | Standard Error | t-Statistic | P-Value |
| --- | --- | --- | --- | --- | --- |
| 50–100 | α | 0.030 | 0.0024 | 12.34 | $9.84 \times 10^{-22}$ |
| | β | 0.28 | 0.026 | 10.68 | $3.54 \times 10^{-18}$ |
| | γ | 0.045 | 0.0023 | 19.43 | $1.47 \times 10^{-35}$ |
| 200–320 | α | 0.19 | 0.043 | 4.48 | 0.000012 |
| | β | 0.14 | 0.012 | 12.05 | $1.87 \times 10^{-26}$ |
| | γ | 0.15 | 0.012 | 12.82 | $5.46 \times 10^{-29}$ |

**Table 4**
Parameters for Planck Blackbody.

| Parameter | Estimate | Standard Error | t-Statistic | P-Value |
| --- | --- | --- | --- | --- |
| $C_1$ | 0.070 | 0.00024 | 288.63 | $8.17 \times 10^{-232}$ |
| $C_2$ | $2.78 \times 10^{-11}$ | $8.85 \times 10^{-13}$ | 31.44 | $5.62 \times 10^{-73}$ |

This study's results on various modified disease spread models illustrate the importance of social distancing and its effects on the rise of infections during a pandemic. The ability of a population to adhere to social parameters set by the government can greatly influence and control the spread of an infection. The *m* factor presents a good representation of adherence to social parameters; however, it is important to note that many other factors can be introduced to better reflect these anthropological variables which are subject to change.

*4.2. SIR model inflection points*

An important aspect of disease modelling is understanding the peak of infection. An inflection point in the curve would suggest the peak of infection has been reached which may not be visible using the variable, time (t), in the earlier stages of the spread. For this reason, it is important to be able to use the *s, i,* and *r* variables independently to derive the inflection point without depending on time (t) as a variable.

In this case, the condition to determine an inflection point are as follows, recalling that Eq. (5) states $\frac{ds}{dt} = -\beta s i$:

$$s'' = \frac{d^2 s}{dt^2} \tag{34a}$$

For an inflection point to occur, $s'' = 0$ and $si$ are constant

$$\frac{ds}{dt} i + \frac{di}{dt} s = 0 \tag{34b}$$

$$\frac{di}{dt} = \beta s i - \gamma i \tag{34c}$$

Eq. (34c) can be rewritten as:

$$\frac{ds}{dt} i = -s \frac{di}{dt} = -s i (\beta s - \gamma) \tag{34d}$$

Factoring out *i*, we have a simplification,

$$\frac{ds}{s} = -(\beta s - \gamma) dt \tag{34e}$$

or

$$\frac{ds}{s(\beta s - \gamma)} = -dt \rightarrow \int \left( \frac{1}{s} - \frac{\beta}{\beta s - \gamma} \right) ds = \gamma \int dt \tag{34f}$$

$$\log(s) - \log(\beta s - \gamma) = \gamma t \rightarrow \log\left(\frac{s}{\beta s - \gamma}\right) = \gamma t \tag{34g}$$

$$t_{inflection} = \frac{\log(s) - \log(\beta s - \gamma)}{\gamma} \tag{34h}$$

*4.3. Planck–Like blackbody distribution and infectivity*

When analyzing several SIR models of disease, it was observed that the infection curve can resemble the Planck-like Blackbody function curves. Planck's Blackbody Distribution is known to have two dependents: wavelength and temperature. While the SIR models illustrate singular dependence, the SIRmp model introduces a second factor that, much like how the temperature factor in a blackbody affects the peak of the intensity, can affect the rise of infections according to time and change the peak's position on the graph.

A blackbody is a physical phenomenon that absorbs all incidence of radiation while emitting a continuous spectrum dependent on its thermal conditions. The higher the temperature of the blackbody, the higher the peak of re-emission intensity at a lower wavelength [39].

This can be compared to the infection curve in the SIRmp model. The mixing factor, referred to as the *m* factor, is much like the temperature factor of the blackbody. If the population of a country is akin to the blackbody, a high *m* factor value of a population will allow for a maximum peak of infection to occur earlier during the pandemic. This is similar to the temperature variable in a blackbody, which can induce a maximum peak of the intensity at a lower wavelength. This allows the m factor to present a measure of how much a population obeys social distancing measures provided by the government.

This comparison presents a good approach as to how the infection rate of a virus can depend on both time and compliance attributes of a population.

**5. Conclusions**

The equations used in the SIR model were time dependent [1]. This study examined not only the time dependent equations but also derived the different variable relationships to one another. Specifically, this study derived the equation for the number of infected cases depending on the number of susceptible individuals, which in turn was found with respect to removed individuals. These equations allow for the study of infection in relation to transmission. That is, using these models, one can now mathemat-ically study the relationship between infected, susceptible and removed individuals in epidemic models.

With the SEIRρqr model, this study wanted to examine the impact of protective procedures on reducing disease spread. The equations were modelled to account for the effect of social distancing on the SEIRρqr model - particularly, the exposed and infected groups by the variable $\rho$ [36]. While the feasibility of complete adherence is difficult, these results support ideas of protective measures in reducing exposure - therefore, infection - of disease, particularly with the encouraging emergence of vaccines such as Pfizer, Moderna, and Johnson & Johnson etc. [41–43]. The models discussed in this study have a good range of variability and applicability – but they are not perfect. It is important to note that these models assume ideal conditions so they may not truly reflect the true situation – for example, these models do not consider the implication of asymptomatic cases which may not be identified [6]. The reality of rapidly changing data and the need to update information based on the new data emphasizes the importance of dynamic rather than static modelling [3]. Moreover, the capability of model updating on a regular basis needs to be developed [3]. Another major consideration of using these mathematical models is the problem of model calibration. In order to determine accurate parameters that reflect measured data, the inverse modelling problem must be solved [3,6,8,19,38]. That is, one must devise a method to accurately estimate the parameters of these mathematical models to reproduce measurable data reflective of the true situations [3,6,8,19,38]. As this study focused on ideal conditions, the inverse modelling problem was not addressed.

To our knowledge, no other study has examined COVID-19 transmission with respect to the SIR model using specific variable related derivations, the SEIRρqr model with focus on impact of social distancing and the similarities of the infection curves to Planck-like blackbody functions. This study presented several mathematical approaches for the modelling of disease transmission using methodologies ranging from the SIR model to the SEIRρqr model, and simulations by the Planck blackbody function. Specifically, it demonstrated practical applications of these models by comparing their results fitted onto the Canadian COVID-19 cases data. Through the predicted values from each model, meaningful inferences about the behaviour and trajectory of the COVID-19 pandemic were drawn.

The results of this study can be used to better understand - or help confirm - the trends of COVID-19 transmission in a Canadian context. Further studies can use this data to further investigate the efficacy of using these mathematical models in extrapolating COVID-19 transmission trends – including trends of new dangerous variants such as Delta and Omicron.

## CRediT authorship contribution statement

**R. Jayatilaka:** Validation, Investigation, Data curation, Visualization, Project administration. **R. Patel:** Validation, Investigation, Data curation, Visualization, Project administration. **M. Brar:** Validation, Investigation, Data curation, Visualization, Project administration. **Y. Tang:** Methodology, Conceptualization, Software, Formal analysis, Investigation, Data curation, Visualization, Project administration. **N.M. Jisrawi:** Methodology, Conceptualization, Formal analysis, Visualization. **F. Chishtie:** Methodology, Conceptualization, Software, Validation, Formal analysis, Investigation, Data curation, Visualization. **J. Drozd:** Methodology, Conceptualization, Software, Validation, Formal analysis, Investigation, Data curation, Visualization. **S.R. Valluri:** Methodology, Conceptualization, Formal analysis, Visualization, Supervision, Project administration.

## Declaration of Competing Interest

The authors declare that they have no known competing financial interests or personal relationships that could have appeared to influence the work reported in this paper.

## Acknowledgement

The authors would like to thank Ken Roberts for his inspiring discussions that have resulted in an improved version of our manuscript. The authors would also like to thank Frank Li for his help in establishing some early coding and stimulating discussion. The authors would also like to thank Purnima Sainani and Rishi Patel for their substantial help in editing the manuscript. We thank the anonymous reviewers for their valuable suggestions and insightful comments which have greatly improved our manuscript.

## References

[1] M. Batista. *Classical deterministic contagious epidemic models without vital dynamics*. 2020. url: https://www.researchgate.net/publication/341250851_Classical_deterministic_contagious_epidemic_models_without_vital_dynamics

[2] Blackbody Radiation. In: *National Radio Astronomy Observatory* (2020). url: https://www.cv.nrao.edu/course/astr534/BlackBodyRad.html

[3] D. Calvetti, A.P. Hoover, J. Rose, E. Somersalo, Network Models for Understanding, Predicting, and Managing the Coronavirus Disease COVID-19, Front. Phys. 8 (2020), https://doi.org/10.3389/fphy.2020.00261/full.

[4] Public Health Agency of Canada. *Coronavirus Disease (COVID-19)*. Aug 2020. url: https://www.canada.ca/en/public-health/services/diseases/coronavirus-disease-covid-19.html..

[5] J.M. Carcione et al., A Simulation of a COVID-19 Epidemic Based on a Deterministic SEIR Model, Front. Public Health 8 (2020), https://doi.org/10.3389/fpubh.2020.00230.

[6] D. Celentano, Gordis Epidemiology, Elsevier, 2019.

[7] S. Chatterjee et al., SEIRD model to study the asymptomatic growth during COVID-19 pandemic in India, Indian J. Phys. 2015 (2020), https://doi.org/10.1007/s12648-020-01928-8.

[8] Y.C. Chen et al. A Time-dependent SIR model for COVID-19 with Undetectable Infected Persons, in: IEEE Transactions on Network Science and Engineering (2020). arXiv: 2003.00122[q-bio.PE]

[9] A. Comunian, R. Gaburro, M. Giudici. Inversion of a SIR-based model: A critical analysis about the application to COVID-19 epidemic, in: Elsevier Public Health Emergency Collection, Physica D: Nonlinear Phenomena (2020). doi: 10.1016/j.physd.2020.132674. url: https://www.ncbi.nlm.nih.gov/pmc/articles/PMC7419377/.

[10] R.M. Corless, G.H. Gonnet, D.E.G. Hare, D.J. Jeffrey, D.E. Knuth, On the Lambert W function, Adv. Computational Math. 5 (1) (1996) 329–359, https://doi.org/10.1007/BF02124750.

[11] P. Erdös, A. Rényi, On the evolution of random graphs, Publ. Math. Inst. Hungary. Acad. Sci. 5 (1961) 17–61.

[12] A. Godio, F. Pace, A. Vergnano, SEIR Modeling of the Italian Epidemic of SARS-CoV-2 Using Computational Swarm Intelligence, Int. J. Environ. Res. Public Health 17 (10) (2020) 3535, https://doi.org/10.3390/ijerph17103535.

[13] A. Grifoni et al., Targets of T Cell Responses to SARS-CoV-2 Coronavirus in Humans with COVID-19 Disease and Unexposed Individuals, Cell (2020), https://doi.org/10.1016/j.cell.2020.05.015.

[14] R. Jeffrey, D. Jeffrey. The Lambert W Function, in: The Princeton Companion to Applied Mathematics (2015). doi: 10.1515/9781400874477. url: https://xueyuechuan.me/files/The%20Princeton%20Companion%20to%20Applied%20Mathematics.pdf.

[15] B. Jester, T. Uyeki, D. Jernigan, Readiness for Responding to a Severe Pandemic 100 Years After 1918, Am. J. Epidemiol. (2018), https://doi.org/10.1093/aje/kwy165.

[16] W.O. Kermack, A.G. McKendrick, A Contribution to the Mathematical Theory of Epidemics, Roy. Soc. (1927) 700–721, https://doi.org/10.1098/rspa.1927.0118.

[17] J. Kranz, Comparison of Epidemics, Ann. Rev. Phytopathol. 12 (1) (1974) 355–374, https://doi.org/10.1146/annurev.py.12.090174.002035.

[18] H.G. Landau, On some problems of random nets, Bull. Math. Biophys. 14 (2) (1952) 203–212, https://doi.org/10.1007/bf02477719.

[19] G. Li, Y. Fan, Y. Lai, T. Han, Z. Li, P. Zhou, P. Pan, W. Wang, D. Hu, X. Liu, Q. Zhang, J. Wu, Coronavirus infections and immune responses, In: J. Med. Virol. 92 (4) (2020) 424–432, https://doi.org/10.1002/jmv.v92.410.1002/jmv.25685.

[20] T. Marinov, R. Marinova, Dynamics of COVID-19 Using Inverse Problem for Coefficient Identification in SIR Epidemic Models, Chaos, Solitons Fractals X (2020), https://doi.org/10.1016/j.csfx.2020.100041.

[21] L. Matrajt, T. Leung, Evaluating the Effectiveness of Social Distancing Interventions to Delay or Flatten the Epidemic Curve of Coronavirus Disease, Emerging Infectious Dis. 26 (8) (2020) 1740–1748.

[22] C. Mills, J. Robins, M. Lipsitch, Transmissibility of 1918 pandemic influenza, Nature 432 (7019) (2004) 904–906, https://doi.org/10.1038/nature03063.
[23] J. Moehlis. An SEIR Model. 2002. url: https://sites.me.ucsb.edu/ ~moehlis/APC514/tutorials/tutorial_seasonal/node4.html.
[24] C Nave. Blackbody Radiation. HyperPhysics (n.d.). url: http://hyperphysics.phy-astr.gsu.edu/hbase/mod6.html.
[25] M. Nikolaou Using Feedback on Symptomatic Infections to Contain the Coronavirus Epidemic: Insight from a SPIR Model 2020 10.1101/2020.04.14.20065698
[26] Wolfram Research. ParametricNDSolve, Wolfram Language function. 2012. url: https://reference.wolfram.com/language/ref/ParametricNDSolve. Html.
[27] K. Roberts. Unpublished raw data of Ken Roberts regarding SEIRm simulations of Ontario COVID-19 cases. 2020.
[28] M. Rojas, Y. Rodríguez, D.M. Monsalve, Y. Acosta-Ampudia, B. Camacho, J.E. Gallo, A. Rojas-Villarraga, C. Ramírez-Santana, J.C. Díaz-Coronado, R. Manrique, R.D. Mantilla, Y. Shoenfeld, J.-M. Anaya, Convalescent plasma in Covid-19: Possible mechanisms of action, Autoimmunity Rev. 19 (7) (2020) 102554, https://doi.org/10.1016/j.autrev.2020.102554.
[29] H.A. Rothan, S.N. Byrareddy, The epidemiology and pathogenesis of coronavirus disease (COVID-19) outbreak, J. Autoimmunity 109 (2020) 102433, https://doi.org/10.1016/j.jaut.2020.102433.
[30] D. Smith, L. Moore. The SIR Model for Spread of Disease - The Differential Equation Model. Mathematical Association of America (2004). url: https://www.maa.org/press/periodicals/loci/joma/the-sir-model-for-spread-of-disease-the-differential-equationmodel.
[31] R. Solomonoff, A. Rapoport, Connectivity of random nets, Bull. Math. Biophys. 13 (2) (1951) 107–117, https://doi.org/10.1007/bf02478357.
[32] J.K. Taubenberger, D.M. Morens, 1918 Influenza: the Mother of All Pandemics, Emerging Infectious Dis. 12 (1) (2006) 15–22, https://doi.org/10.3201/eid1209.05-0979.
[33] S.R. Valluri, D. Jeffrey, R. Corless, Some applications of the Lambert W function to physics, Can. J. Phys. 78 (9) (2000) 823–831, https://doi.org/10.1139/p00-065.
[34] S.R. Valluri, K. Roberts, Notes on Solutions of SIR-Type Epidemic Models (2020).
[35] E Weisstein. *Kermack-McKendrick Model*. url: https://mathworld.wolfram.com/Kermack-McKendrickModel.html.
[36] E. Weisstein. *SIR Model*. url: https://mathworld.wolfram.com/SIRModel.html.
[37] P.J. Witbooi, G.E. Muller, G.J. Van Schalkwyk, Vaccination Control in a Stochastic SVIR Epidemic Model, Comput. Math. Methods Med. 2015 (2015) 1–9, https://doi.org/10.1155/2015/271654.
[38] R.M. Corless, D.J. Jeffrey, The Wright ω function. In *Artificial intelligence, automated reasoning, and symbolic computation*, Springer, Berlin, Heidelberg, 2002, pp. 76–89.
[39] R. Schlickeiser, M. Kröger, Epidemics forecast from SIR-modeling, verification and calculated effects of lockdown and lifting of interventions, Front. Phys. 8 (2021) 593421.
[40] S. Thornton, A. Rex. Modern Physics for Scientists and Engineers 1(4) (2013) 96-97.
[41] R. Schlickeiser, M. Kröger, Analytical solution of the SIR-model for the temporal evolution of epidemics: Part B. Semi-time case, J. Phys. A: Math. Theoretical 54 (17) (2021) 175601, https://doi.org/10.1088/1751-8121/abed66.
[42] K. Bok, S. Sitar, B.S. Graham, J.R. Mascola, Accelerated COVID-19 vaccine development: milestones, lessons, and prospects, Immunity 54 (8) (2021) 1636–1651.
[43] Y. Li, R. Tenchov, J. Smoot, C. Liu, S. Watkins, Q. Zhou, A comprehensive review of the global efforts on COVID-19 vaccine development, ACS Central Sci. 7 (4) (2021) 512–533.